**Phase transition and polar cluster behavior above Curie temperature in ferroelectric BaTi$_{0.8}$Zr$_{0.2}$O$_3$**

Oktay Aktas,[1,*] Francisco Javier Romero,[2] Zhengwang He,[1] Gan Linyu,[1] Xiangdong Ding,[1] José-María Martín-Olalla,[2] María-Carmen Gallardo,[2] and Turab Lookman[3]

[1]*State Key Laboratory for Mechanical Behavior of Materials & Materials Science and Engineering, Xi'an Jiaotong University, Xi'an 710049, China*
[2]*Departamento de Física de la Materia Condensada, ICMSE-CSIC, Universidad de Sevilla, Apartado 1065, 41080 Sevilla, Spain*
[3]*AiMaterials Research LLC, Santa Fe, NM 87501, USA*

[*]Author to whom correspondence should be addressed: Oktay Aktas, oktayaktas@xjtu.edu.cn.

**Abstract**

*We study the phase transition behavior of the ferroelectric BaTi$_{0.8}$Zr$_{0.2}$O$_3$ in the paraelectric region. The temperature dependencies of thermal, polar, elastic and dielectric properties indicate the presence of local structures above the paraelectric-ferroelectric transition temperature $T_c$ = 292 K. The non-zero remnant polarization is measured up to a characteristic temperature T\* ~350 K, which coincides with the temperature where the dielectric constant deviates from Curie-Weiss law. Resonant Piezoelectric Spectroscopy shows that DC field-cooling above $T_c$ using fields smaller than the coercive field leads to an elastic response and remnant piezoelectricity below T\*, which likely corresponds to the coherence temperature associated with polar nanostructures in ferroelectrics. The observed remnant effect is attributed to the re-orientation of polar nanostructures above $T_c$.*

Paraelectric phases have long been considered to have nanoscale structures that are polar and dynamic,.[1,2] However, recent transmission electron microscopy and Raman scattering experiments[3] on $BaTiO_3$ have shown that such structures with size 4 nm-6 nm can be static at least over a large temperature range (~100°C) and over time scales of hundreds of seconds.[3] In relaxor ferroelectric lead magnesium niobate ($PbMg_{1/3}Nb_{2/3}O_3$, PMN), domain structures were previously visualized at room temperature,[4] which is well above the freezing temperature ($T_f$ ~220 K). This is consistent with the results of simulations on relaxor ferroelectrics that indicate that the paraelectric phase consists of many such interacting polar nanoclusters without the parent phase.[5] These simulations indicate co-existence of static and dynamic clusters below the coherence temperature. Calculations on $BaTiO_3$ have suggested that the paraelectric phase contains regions with Ti-off centering that lead to a global polarization, that was absent in cubic $BaZrO_3$.[6,7] Moreover, pyroelectric measurements on PMN have suggested that such structures can be reoriented by the application of an electric field.[8] The existence of local polar nanostructures is important not only for understanding relaxors and ferroelectrics but also their applications in nanotechnology.[9-14]

Although relaxor PMN shows evidence of poling of the microstructure,[8] such investigations on ferroelectrics are rare, despite the observation of static-like clusters.[3] In this Letter, we study the phase transition behavior of ferroelectric $BaTi_{0.8}Zr_{0.2}O_3$ to suggest the reorientation of polar nanostructures under applied fields. We show that remnant piezoelectricity exists in the paraelectric phase and does not dissipate for many days.

$BaTi_{1-x}Zr_xO_3$ (BZT) solid solutions undergo a ferroelectric transition for x<0.25 but lead to relaxor behavior for higher concentrations x.[15-18] Analogous to $BaTiO_3$, BZT solid solutions with x<0.15 undergo a paraelectric-ferroelectric-ferroelectric-ferroelectric phase transition sequence corresponding to cubic-tetragonal-orthorhombic-rhombohedral structural changes. The paraelectric-ferroelectric transition in the range 0.15<x<0.25 leads to a single ferroelectric transition (likely rhombohedral R3m structure). Compositions x>0.25 show relaxor behavior. BZT solid solutions and their derivatives are technologically important for various applications.[10, 18-21]

Ceramic disks, hereafter the samples, with a diameter of 1.2 cm and thickness of ~10 mm and composition 0.80 ($BaTiO_3$)-0.20 ($BaZrO_3$) were fabricated by conventional solid-state reaction method with starting chemicals $BaTiO_3$ (99.9%) and $BaZrO_3$ (99%). The calcination was performed at 1350°C and sintering was done at 1450°C in air. The sample was preliminarily characterized by x-ray diffraction and no secondary phases were detected (Fig. S1). Hysteresis loops of polarization versus electric field [P(E)] were measured by a driving electric field of 20 kV/cm at 5 Hz during cooling with a Radiant Precision Premier II tester (P-HVi4K), Radiant Precision 4KV HVI(P-PM2) and High voltage amplifier (Trek 609E-6). Dielectric constant of the sample was measured as a function of temperature by using a LCR meter in a nitrogen-gas-cooled furnace on cooling at a rate of 2 K/min.

One disk was cut into a parallelepiped with dimensions ~7.2 mm × ~8 mm × 1 mm. Resonant Ultrasound Spectroscopy (RUS)[22-24] measurements were then conducted using lead zirconate titanate transducers with a ten-minute settling time by using a nitrogen-gas-cooled Furnace (Suns Electronic Systems, model EC1X). A function generator/lock-in amplifier unit (Zurich Instruments, model HF2LI) was used for the generation and detection of mechanical resonances in RUS measurements.

Another disk was used to measure the pyroelectric current at zero electric field, after poling it with E = 3.1 kV/cm at 243 K. The polarization was determined by integrating the pyroelectric current. To correlate the polarization with the excess entropy, specific heat measurements were conducted using the thermal relaxation method at a Physical Properties Measurement System (PPMS, Quantum Design at CITIUS, University of Seville / Spain). From the disk, a parallelepiped measuring 3.0 mm × 3.0 mm × 0.9 mm and 31.0 mg in mass was cut. The accuracy and precision of this technique have been reported previously.[25]

We used resonant piezoelectric spectroscopy (RPS)[24,26,27] to investigate the electrical switching behavior before, under and after DC fields in the paraelectric phase of BZT20. RPS is the electrical analogue of RUS. It was previously used to probe miniscule, symmetry-disallowed piezoelectricity in nominally centrosymmetric materials[26] and to quantitatively measure the temperature evolution of the piezoelectric coefficient $d_{33}$ upon in-situ electrical poling of ferroelectrics.[27] For RPS measurements under DC bias, the alternating voltage from the function generator was fed into a KROHN-HITE wideband amplifier (model 7602M).

The phase transition behavior of BZT20, investigated by specific heat, RUS, and dielectric constant measurements, is presented in Fig. 1. The temperature dependence of the specific heat collected on heating and cooling shows a maximum at $T_c$ = 292 K. Any thermal hysteresis that may exist is not well-resolved, indicating nearly continuous behavior of the phase transition from the paraelectric phase with Pm-3m symmetry to the ferroelectric phase with rhombohedral R3m symmetry. In Fig. 1(b), the squared frequencies of two mechanical resonances collected by RUS are plotted against temperature (see Fig. S2 for RUS spectra collected as a function of temperature). The temperature of the sharp minimum in the squared frequencies coincides with $T_c$ determined from the specific heat (within 1 K) and corresponds to the transition temperature, as shown in other ferroelectric perovskites.[23,24] Also, RUS heating and cooling data show little thermal hysteresis (see inset of Fig 1(b)) in agreement with the specific heat measurements. Therefore, the first order nature of the transition is very weak, if it exists. Moreover, despite a factor of ~5 in their resonance frequencies, both mechanical resonances have the same minimum and show no frequency dispersion. Similarly, the dielectric constant peak is at the same temperature (295 K) for frequencies between 100 Hz and 1 MHz. This temperature is slightly higher than $T_c$ measured by the elastic modulus and the specific heat. This difference cannot be accounted for by potential errors in temperature calibration. Therefore, BZT20 shows a weakly diffuse transition that is sometimes considered to be an intermediary between a relaxor and a ferroelectric transition. In agreement with this, the fit of the dielectric constant by the modified Curie-Weiss law leads to an exponent of 1.7, which would be 1 for classical ferroelectrics and 2 for relaxors (Fig. S3). Similar intermediate states were previously found in lead-based materials.[29]

The squared frequency of a mechanical resonance measured by RUS is proportional to an effective elastic modulus dominated by shearing. Thus, in Fig. 1(b), the reduction of the squared frequencies at $T_c$ with respect to the highest measuring temperature corresponds to a softening of 31% in the elastic modulus. According to Landau theory, the cubic-rhombohedral structural transition (Pm-3m to R3m) occurring at $T_c$ is improper ferroelastic, which would normally lead to a step-change in the elastic modulus at $T_c$.[23] The observed nonlinear elastic softening in the paraelectric phase is instead explained by the nanostructure and its change with temperature.[23,24,28,30] In Fig. 1(d), we explore the scaling relationship between the excess entropy $\Delta S$ and the squared polarization $P^2$. The excess entropy was derived by integration of $\Delta c/T$, where $\Delta c$ represents the excess specific heat relative to the baseline shown in Fig. 1(a) (see also Fig. S4). The squared polarization was calculated by integration of the pyroelectric current measured as a function of temperature (Fig. S5). $\Delta S$ and $P^2$ linearly scale for temperatures below 299 K, in line with a single order parameter with a dominating temperature dependent linear term in Landau theory.[31] However, above $T_c$ over a temperature range of 50 K, differences in the non-linear region above $T_c$ are easily observed, which means lack of coupling between the two magnitudes. Indeed, the transition to the ferroelectric phase contains ferroelectric displacements and their coupling with strain[23] and possibly gradient coupling terms.[32,33,34] Such coupling can contribute towards the presence of local polar structures in the paraelectric phase. Thus, in addition to the elastic modulus, the non-linear scaling of $\Delta S$ and $P^2$ is also indicative of local polar structures in the paraelectric phase.

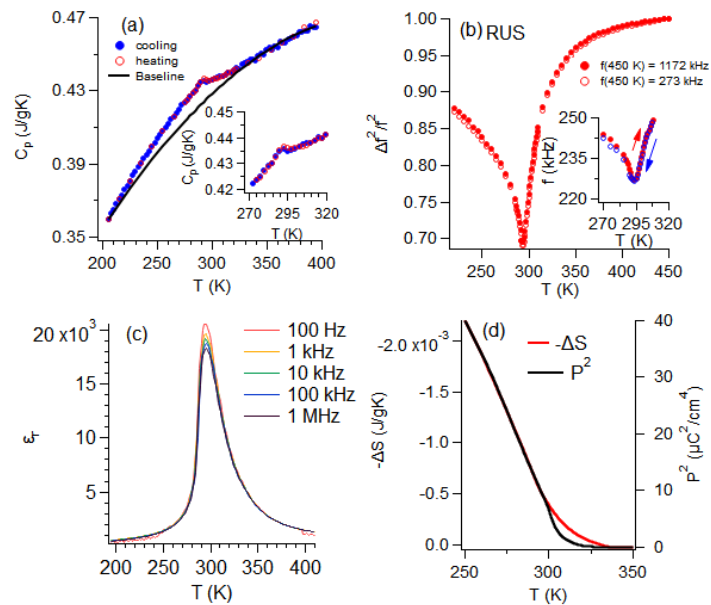

**Fig. 1** Phase transition behavior and scaling of the excess entropy and the squared polarization (a) Temperature dependence of the specific heat. The inset shows the anomaly in a narrower temperature range. (b) Temperature dependence of the squared mechanical resonance frequencies, proportional to an effective elastic modulus. The inset shows the variation of the resonance frequency obtained on cooling and heating. (c) Temperature dependence of the dielectric constant. (d) The temperature evolution of the excess entropy and the polarization squared, which shows departure from linear scaling in the paraelectric phase.

We next examine the ferroelectric switching behaviour through hysteresis measurements of polarization vs. electric field collected on cooling (Fig. 2(a)). Hysteresis loops typical of a ferroelectric phase are observed below $T_c$ (solid lines). Loops reminiscent of ferroelectric switching are observed even above $T_c$ (dashed lines), although they are relatively slim and gradually narrow down with increasing temperature. Above T* = 350 K-360 K, the loops are primarily determined by leakage currents (yellow dotted lines, 378 K). The overall temperature dependence of $P_r$ (Fig. 2(b)) is rather continuous, with non-zero remnant polarization detected up to T* ~ 350K-360 K. The temperature T* roughly coincides with the temperature where deviation from Curie-Weiss law is observed via the inverse dielectric constant $1/\varepsilon_r$ (right axis). In reference to extensive studies on 0.70 ($PbMg_{1/3}Nb_{2/3}$)-0.30 ($PbTiO_3$),[35, 36] T* can be conjectured to be the coherence temperature. It is characterized by the development of static polar nanostructures,[5, 35, 36, 37] which can contribute to the persistence of remnant polarization between $T_c$ and T*.

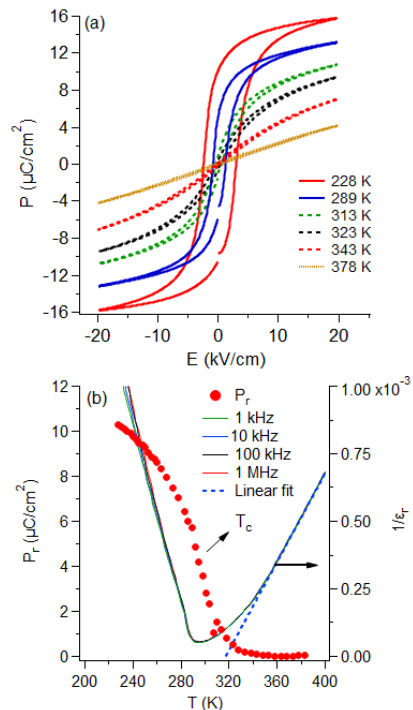

**Fig. 2.** Remnant Polarization as a function of temperature. (a) Hysteresis curves of polarization vs. electric field (b) Temperature dependence of remnant polarization $P_r$ and $1/\varepsilon_r$.

We further explore the electrical switching behavior in the paraelectric phase by RPS measurements conducted under DC field-cooling on samples annealed at various temperatures. Like RUS, the square of the mechanical resonance frequencies measured by RPS is proportional to the elastic modulus. Additionally, the area under the RPS spectrum provides a quantitative measure of the piezoelectric coefficient, $d_{33}$,[27] as a function of temperature and allows in-situ monitoring under applied electric fields.[26] This has previously been used to measure symmetry-disallowed, miniscule piezoelectricity in paraelectric phases of ferroelectrics, relaxor ferroelectrics, ferroelastics, and compounds containing defects.[27] Here, RPS is used to compare the effects of DC field-cooling on the symmetry-disallowed piezoelectricity in the paraelectric phase. The annealing temperatures were selected to be 323 K, 343 K, 380 K, and 460 K. These temperatures correspond to temperature ranges below T* ~350-360 K, between T* and the projected Burns temperature $T_B$ ~450 K,[16,17] and above $T_B$. The applied field was E = 0.65 kV/cm, which is a fourth of $E_c$ at 228 K. Segments of RPS spectra obtained at 323 K are shown in Fig. 3. They were collected under the following conditions: 1) The sample was annealed at 460 K for two days and then zero-field cooled to 323 K; 2) The sample was annealed at 460 K for two days and the spectrum was collected at 323 K 66 hours after it was field-cooled to 323 K; 3) The sample was annealed at 343 K (below T*~350-360 K) for two days and the spectrum was collected 44 hours after the sample was cooled to 323 K; 4) The sample was annealed at 380 K for one day and the spectrum was collected 10 minutes after the sample was cooled to 323 K.

To obtain a measure of the magnitude of the piezoelectric effect, we calculated RPS spectrum areas by integration as plotted in Fig. 3(b). The piezoelectric effect shows a 3-fold increase upon field cooling compared to before field cooling. Changes in mechanical resonance frequencies and linewidths of peaks correspond to elastic hardening and increased acoustic attenuation. Even after annealing ~2 days at 343 K (below T* determined by deviation from the Curie-Weiss law in Fig. 2(b)), the piezoelectric coefficient of the sample is slightly larger than that before poling. However, annealing above T* (at 380 K) for 10 minutes leads to a similar piezoelectric coefficient as before field-cooling, indicating a rapid depolarization process. Thus, T* is identified as the temperature below which remnant piezoelectricity persists.

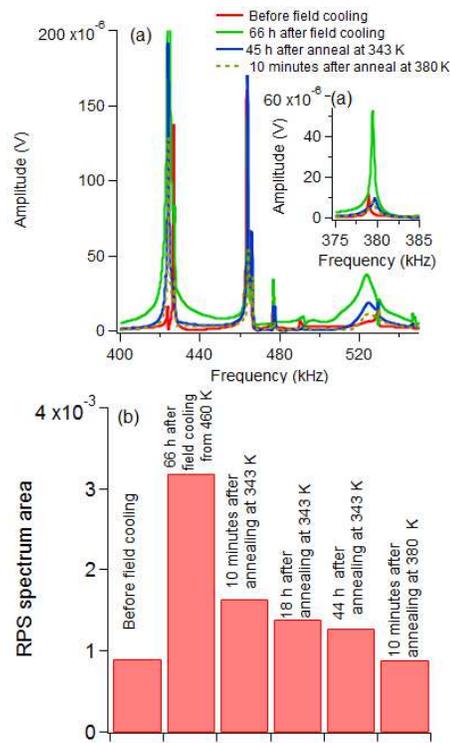

**Fig. 3.** Electrical switching behavior with E = 0.65 kV/cm <1/3 $E_c$ (260 K) due to field cooling to 323 K and under different annealing. (a) Segments of RPS spectra collected after different annealing and field-cooling procedures (also see text) (b) RPS spectrum areas for comparing effects of high temperature annealing and field-cooling.

A field induced transition for the remnant piezoelectricity is unlikely for several reasons. The applied field was much lower than $E_c$, whereas field-induced transitions in the paraelectric phase would typically require a field higher than $E_c$. Moreover, field-induced transitions are reversible except in several compounds in which a relaxor ferroelectric-ferroelectric transition (RFE-FE) occurs around a temperature where the dielectric constant shows a hump.[38] The maximum of the dielectric constant is at a much higher temperature. In contrast, the remnant piezoelectricity in BZT20 is observed ~30 K above the temperature of the dielectric maximum, which is close to $T_c$ (Fig. 1).

We attribute the remnant piezoelectricity below T* to the reorientation of the polar nanostructures. First of all, the observed elastic variations in Fig. 1 and Fig. 3(b) show aspects of glass-like behavior proposed in parent phases of ferroelectrics and in ferroelastic materials.[39-41] The time dependence of the remnant piezoelectricity measured as a function of time after field-cooling can be fitted to a stretched exponential, typical of glassy systems (Fig. S6).[39,42,443] Moreover, the piezoelectric coefficient of the paraelectric phase is three orders of magnitude larger than the values observed in compounds for which no evidence of polar nanostructures is reported and which are instead known to contain defects.[27] In addition, the observed piezoelectricity is enhanced as the temperature approaches $T_c$. The magnitude of the piezoelectric effect is ~20 times larger at 323 K than at 391 K (Fig. S7(a)). The pronounced difference in magnitudes reflects increased coherence of polar nanostructures on cooling. Complementary measurements of RPS under a DC bias conducted below and above T* are shown in Fig. 4 (and Fig. S7). Under a DC bias, the observed piezoelectric effect is much larger below T* than above T*. Also, the time dependence of the piezoelectric effect below and above T* suggests slower dynamics below T*. At 118°C the piezoelectric effect is nearly stable after 80 minutes, whereas much longer times are needed for 50°C. All these observations, along with the observation of remnant piezoelectricity at least up to T*, point towards a dominant role of static-like polar nanostructures in the remnant piezoelectric effect.

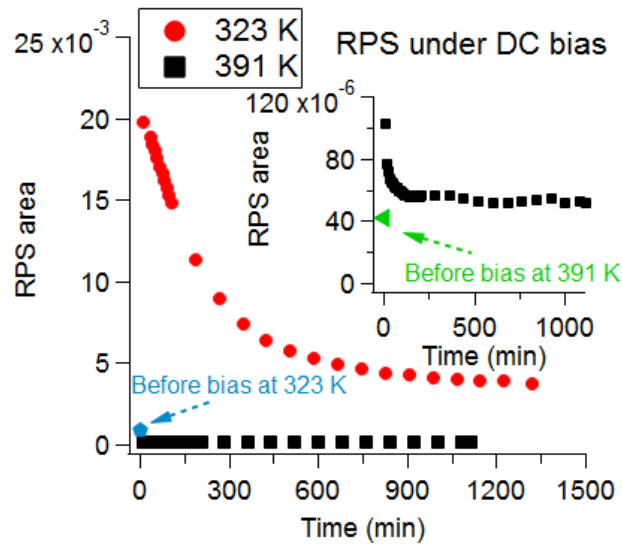

**Fig. 4.** Time dependence of the magnitude of the piezoelectric effect under DC bias with E = 650 V/cm (in the form of RPS spectrum area). The blue pentagon and green triangle correspond to the spectrum areas before applying the bias.

While pyroelectric measurements[44] show evidence of switchable polar nanostructures in relaxor ferroelectric PMN,[8] the results presented here suggest that even in the case of a weakly diffused ferroelectric transition, such polar structures may remain electrically switched for days above $T_c$. For further clarification of the evolution of polar nanostructures in ferroelectrics and relaxors, it is crucial to test whether remnant piezoelectricity is permanent. The response of polar nanostructures to electric fields suggested here in BZT20 may be relevant for the potential design of materials for various applications, such as tunable capacitance and the electrocaloric effect.

**SUPPLEMENTARY MATERIAL**

See supplementary material for the X-ray diffraction pattern of BZT20, RUS spectra collected as a function of temperature, the fit of the dielectric constant data by the modified Curie-Weiss law, description of relaxational calorimetry, the temperature dependences of the excess specific heat and pyroelectric current, and time dependent RPS measurements.

**ACKNOWLEDGEMENTS**

F. J. R., J.-M. M.-O and M.-. G. acknowledge funding from Universidad de Sevilla (VI PPITU and VII PPITU). O.A. thanks the Natural National Science Foundation of China (NSFC) for financial support (grant no: 51850410520). X.D. thanks NSFC for financial support (grant no: 51931004). O.A. and Z.W.H. thank Lixue Zhang (Xi'an Jiaotong University) for letting them use the ferroelectric workstation in her laboratory and Tianran Zhang (Xi'an Jiaotong University) for her help with the measurements.

1. S.H. Wemple, Phys. Rev. B **2**, 2679 (1970).
2. R. Comes, M. Lambert, A. Guinier, Solid State Commun. **6**, 715 (1968).
3. Bencan, E. Oveisi, S. Hashemizadeh, V. K. Veerapandiyan, T. Hoshina, T. Rojac, M. Deluca, G. Drazic, and D. Damjanovic, Nat. Commun. **12**, 3509 (2021).
4. M. Eremenko, V. Krayzman, A. Bosak, H.Y. Playford, K.W. Chapman, J.C. Woicik, B. Ravel, and I. Levin, Nat. Comm. **10**, 2728 (2019)
5. H. Takenaka, I. Grinberg, S. Liu, A.M. Rappe, Nature **546**, 391 (2017).
6. X.G. Zhao, O. I. Malyi, S.J. Billinge, A. Zunger, Phys. Rev. B **105**, 224108 (2022).
7. J. Očenášek, J. Minár, and J. Alcalá, NPJ Comput. Mater. **9**, 118 (2023).
8. L. M. Riemer, K. Chu, Y. Li, H. Uršič, A. J. Bell, B. Dkhil, and D. Damjanovic, Appl. Phys. Lett. **117**, 102901 (2020).
9. D. D. Viehland and E. K. H. Salje, Adv. Phys. **63**, 267 (2014).
10. X.G. Tang, K.-H. Chew, H.L.W. Chan, Acta Mater. **52**, 5177 (2004)..
11. H. Zhang, H. Giddens, Y. Yue, X. Xue, V. Araullo-Peters, V. Koval, M. Palma, I. Abrahams, H. Yan, Y. Hao, J. Eur. Ceram. Soc. 40, 3996–4003 (2020).
12. L.M. Garten, P. Lam, D. Harris, J.P. Maria, S. Trolier-McKinstry, J. Appl. Phys. **116**, 044104 (2014).
13. J. Li, J. Li, S. Qin, X. Su, L. Qiao, Y. Wang, T. Lookman, and Y. Bai, Phys. Rev. Appl. **11**, 044032 (2019).
14. M. Kumar, G. Sharma, S.D. Kaushik, A. Kumar Singh, S. Kumar, J. Alloys and Comp. **884**, 161067 (2021)
15. D. Hennings, A. Schnell, G. Simon, J. Am. Cer. Soc. 65, 539 (1982).
16. T. Maiti, R. Guo and A. S. Bhalla, J. Am. Ceram. Soc. **91**,1769 (2008).
17. V. V. Shvartsman and D. C. Lupascu, J. Am. Ceram. Soc. **95**, 26 (2012)
18. S. Wada, H. Adachi,H. Kakemoto, H. Chazono, Y. Mizuno, H. Kishi and T. Tsurumi, J. Mater. Res. **17**, 456 (2002)
19. L. Dong, D.S. Stone, R.S. Lakes, J. Appl. Phys. **111**, 0841074 (2012).
20. L. Jin, J. Qiao, L. Wang, L. Hou, R. Jing, J. Pang, L. Zhang, X. Lu, X.Wei, G. Liu, Y. Yan, J. Alloys and Comp. **784**, 931 (2019).
21. N.C.T. Ngo, H. Sugiyama, B.A.K. Sodige, J.P. Wiff, S. Yamanaka, Y. Kim, J. Am. Ceram. Soc. **106**, 201 (2023)

22. A. Migliori and J. Sarrao, *Resonant Ultrasound Spectroscopy: Applications to Physics, Materials Measurements, and Nondestructive Evaluation* (Wiley, New York, 1997).
23. M. A. Carpenter, J. Phys.: Condens. Matter **27**, 263201 (2015).
24. O. Aktas, M. Kangama, G. Linyu, X. Ding, M.A. Carpenter, E.K.H. Salje, , J. Alloys and Comp. **903**, 163857 (2022).
25. J. Lashley, M. Hundley, A. Migliori, J. Sarrao, P. Pagliuso, T. Darling, M. Jaime, J. Cooley, W. Hults, L. Morales, D. Thoma, J. Smith, J. Boerio-Goates, B. Woodfield, G. Stewart, R. Fisher, and N. Phillips, Cryogenics 43, 369 (2003).
26. Z. W. He, O. Aktas, G. Linyu, L.-N. Liu, P. S. da Silva Jr., F. Cordero, X.-M. Chen, X. Ding, and E. K.H. Salje, Journal of Alloys and Compounds **918**, 165783 (2022).
27. O. Aktas, M. Kangama, G. Linyu, G. Catalan , X. Ding, A. Zunger, and E. K. H. Salje, Phys. Rev. Res. **3**, 043221 (2021).
28. G. Linyu, F. J. Romero , V. Franco , J.-M. Martín-Olalla , M. C. Gallardo , E. K. H. Salje , Y. Zhou, and O. Aktas, Appl. Phys.Lett. **115**, 161904 (2019).
29. S. Kustov, J. Miguez Obrero ,X. Wang, D. Damjanovic , and E. K. H. Salje, Phys. Rev. Mater. **6**, 124414 (2022)
30. F. Cordero, F. Trequattrini, P.S. da Silva Jr., M. Venet, O. Aktas, E.K.H. Salje, Phys. Rev. Res. **5**, 013121 (2023).
31. J. M. Martín-Olalla, F. J. Romero, S. Ramos, M. C. Gallardo, J. M. Perez-Mato, and E. K. H. Salje, J. Phys.: Condens. Matter **15**, 2423 (2003).
32. W. Ma and L. E. Cross, Appl. Phys. Lett. **88**, 232902 (2006)
33. S. Conti, S. Muller, A. Poliakovsky, E. K. H. Salje, J. Phys.: Condens. Matter **23**, 142203 (2011).
34. A. N. Morozovska, E. A. Eliseev, M. D. Glinchuk, L.-Q. Chen, and V. Gopalan, Phys. Rev. B **85**, 094107 (2012).
35. J. Narvaez and G. Catalan, Appl. Phys. Lett. **104**, 162903 (2014).
36. Dkhil, P. Gemeiner, A. Al-Barakaty, L. Bellaiche, E. Dul'kin, E. Mojaev, and M. Roth, Phys. Rev. B **80**, 064103 (2009).
37. E. Dul'kin, J. Petzelt, S. Kamba, E. Mojaev, and M. Roth, Appl. Phys. Lett. **97**, 032903 (2010)
38. H. Simons, J.E. Daniels, J. Glaum, A.J. Studer, J.L. Jones, and M. Hoffman, Appl. Phys. Lett. **102**, 062902 (2013)
39. McCloy, J.S. (2019). Spin and Ferroic Glasses. In: Musgraves, J.D., Hu, J., Calvez, L. (eds) Springer Handbook of Glass. Springer Handbooks. Springer, Cham.
40. D. Wang, X. Ke, Y. Wang, J. Gao, Y. Wang, L. Zhang, S. Yang, and X. Ren, Phys. Rev. B **86**, 054120 (2012)
41. E. K. H. Salje, X. Ding, and O. Aktas, Domain glass, Phys. Status Solidi B **251**, 2061 (2014).
42. X. Ding, T. Lookman, Z. Zhao, A. Saxena, J. Sun, and E.K.H. Salje, Physical Review B **87**, 094109 (2013)
43. S.C. Glotzer, N. Jan, T. Lookman, A.B. MacIsaac, and P.H. Poole, Phys. Rev. E **57**, 7350 (1998).
44. S. Hashemizadeh, A. Biancoli, and D. Damjanovic, J. Appl. Phys. **119**, 094105 (2016).